\documentclass[aps,prl,reprint]{revtex4-1}
\usepackage{graphicx}
\usepackage[dvipsnames, svgnames, x11names]{xcolor}  

\usepackage{url}
\usepackage{amsmath}
\usepackage[caption=false,farskip=0pt,labelfont={bf}]{subfig}
\usepackage{dcolumn}
\usepackage{bm}
\usepackage{color}
\usepackage{booktabs} 
\usepackage{mathrsfs}

\usepackage{lineno}

\begin{document}

\preprint{APS/123-QED}

\title{Micromechanical measurements of local plastic events in granular materials}

\author{Jie Zheng}
 \affiliation{School of Physics and Astronomy, Shanghai Jiao Tong University}
\author{Aile Sun}
 \affiliation{School of Physics and Astronomy, Shanghai Jiao Tong University}
\author{Jie Zhang}%
 \email{jiezhang2012@sjtu.edu.cn}
\affiliation{School of Physics and Astronomy, Shanghai Jiao Tong University}%
\affiliation{Institute of Natural Sciences, Shanghai Jiao Tong University}%



\date{\today}

\begin{abstract}
Understanding the relationship between micromechanics and macroscopic plastic deformation is vital for elucidating the deformation mechanism of amorphous solids, such as granular materials. In this study, we directly measure T1 events, which are topological rearrangements of particles, and the associated microscopic stresses in dense packings of photoelastic disks under pure shear. We observe a remarkable similarity between the evolution of the total number of T1 events and the global stress-strain curve. The competition between the birth and death of T1 events establishes a dynamic equilibrium after the yield strain, contributing to the overall plastic behavior. Despite the erratic stress fluctuations of individual T1 events, the local stresses decrease on average once T1 events occur, indicating their soft characteristics. Furthermore, we demonstrate that the microscopic stress fluctuations exhibit a long-range anisotropic spatial correlation similar to the Eschelby character, but with a distinct scaling of $r^{-1.5}$. Interestingly, we also find a striking similarity in the correlation functions of T1 and non-T1 regions. These findings establish a significant connection between macroscopic mechanical behavior and the elementary deformation of T1 events, shedding light on the fundamental understanding of granular materials as amorphous solids with contact-scale plastic deformation.
\begin{description}
\item[DOI] The online like of the paper.
\end{description}
\end{abstract}

\maketitle


\paragraph{Introduction}
Compared to crystalline solids, the microscopic origin governing the deformation of amorphous solids, such as granular materials, colloids, and bulk metallic glasses, is still controversial\cite{barrat2011, RMP, bonn2017yield}. For instance, the nature of ``defects" in amorphous solids\cite{spaepen1977microscopic,spaepen-schall2007structural,argon,argon-kuo1979,argon1982analysis,argon-demkowicz2004high,langer_review,maloney_2004, langer1998, langer2008, Zaccone-2021, Liu-pnas, Liu2015identifying}, the classification of thermodynamic phase transformations during yielding transitions\cite{sastry, procaccia, parisi2017shear, ozawa2018random, sastry2021, Sollich-2022, sastry, sollich1997rheology}, and the microscopic mechanisms governing shear band formation remain subjects of heated debate\cite{YQWANG,band_PRL, procaccia_band, Procaccia-2016, Procaccia-PhysRevE.83.061101, barrat2014spatiotemporal, maloney2004universal, eshelby_pre, maloney_2004}. Therefore, obtaining direct experimental evidence of microscopic plastic events, particularly the associated microscopic mechanical characteristics, is undoubtedly key to resolving these controversies and gaining deeper insights into the deformation mechanisms of amorphous systems.

However, in past experimental approaches, the focus was often on measuring the micro-plastic structure without synchronous micromechanical information, with rare exceptions in flowing emulsions\cite{eric_emulsion}. For instance, in a two-dimensional (2D) bubble system, the elementary plastic event is composed of four particles, known as a T1 event\cite{dennin_2004}. A plastic event occurs when the nearest neighbor and the second nearest neighbor switch. Statistically, it has been found that the number of T1 events is related to the global stress fluctuation of the bubble system\cite{dennin_2004}. However, due to the lack of particle-level mechanical properties in experiments, how these micro-plastic events lead to global stress release remains elusive\cite{foam_eshelby,foam_PRL,foam_prl2}. Previous studies have also found that during the shear process, plastic events are generated and annihilated intermittently, leading to the formation of shear bands in granular systems\cite{granular_prl,granular_prl2}. Yet, the evolving pathway and correlations of these internal plastic events are still controversial due to the lack of direct mechanical information.

Compared to bubble or bulk metallic glasses, densely packed photoelastic particles offer valuable insights into the microscopic stress information of amorphous solids\cite{bob_nature}. Significant advancements have been made in measurement technology and accuracy\cite{photo_1, photo_2}. Previous studies have proposed that plastic deformation in granular materials may arise from force chain buckling\cite{buckling_1} or long-range spatial correlations induced by Eshelby-like ``defects"\cite{eshelby1957,eshelby_pre,eshelby_pre2,granular_prl}, or from macroscopic avalanches triggered by different rotation modes on particle contacts\cite{rolling}. The former focuses on micromechanical behavior, while the latter emphasizes micro-geometric behavior. In this work, we utilize the T1 event as the basic plastic unit, which has been well-studied in bubble, foam and emulsion systems\cite{dennin_2004,foam_prl2,eric_emulsion}. This definition enables us to track and analyze various complex plastic clusters by decomposing them into basic plastic units. By combining the photoelastic technique, we can comprehensively explore the micromechanical characteristics of each event and establish the underlying statistical laws of plastic deformation.

\begin{figure}
\includegraphics[width=8.6cm]{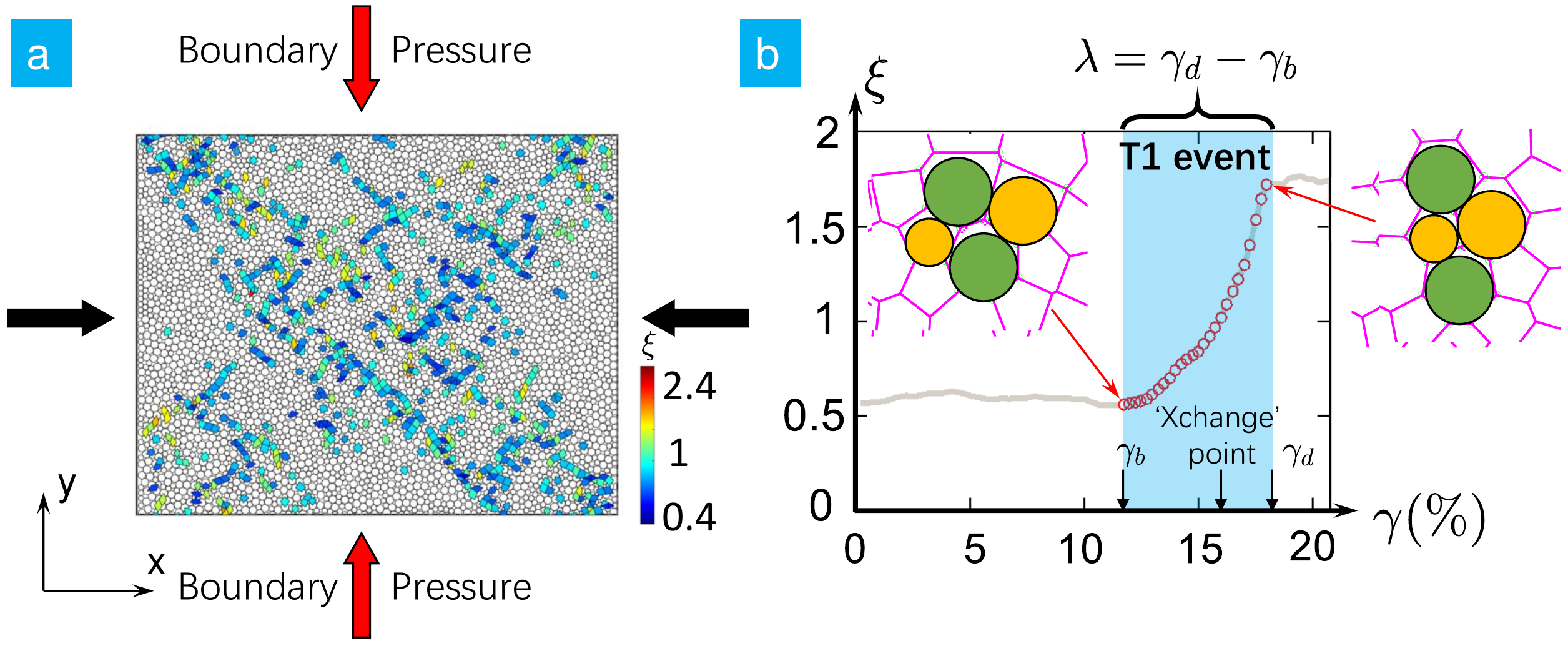}
\caption{(a) Schematic of the two-dimensional photoelastic disk packing subject to pure shear. The bidisperse disks fill up in a rectangular frame subject to pure shear by applying compression along the x direction while allowing free expansion along the y direction that maintains a constant boundary pressure. The individually tracked T1 events are drawn using quadrilaterals connecting the centers of four neighboring particles at the strain of $\gamma=2\gamma_y$. The filled colors represent the instantaneous values of aspect ratios $\xi$ of individual T1 events. (b) Inset is a cartoon plot of a T1 event where two pairs of disks switch neighbors. The main panel depicts the evolution of the aspect ratio $\xi$ of a T1 event versus the strain $\gamma$, from which we define the T1 event’s birth strain, $\gamma_b$, its `Xchange' point, where $\xi=1$, and its death strain, $\gamma_d$. Consequently, the corresponding lifetime of this T1 event $\lambda\equiv\gamma_d-\gamma_b$.}
\label{Fig1} 
\end{figure}

In our bidisperse photoelastic disk experiments, we track T1 events at different strains ($\gamma$) to identify the microscopic plastic deformation unit. We observe a competition between newly generated and annulled T1 events, with a balance point around the yield strain ($\gamma_y$). Interestingly, the total number of T1 events versus strain curve closely resembles the stress-strain curve, indicating a significant connection between macroscopic mechanical behavior and microscopic plastic events. Furthermore, we find that the microscopic stress associated with individual T1 events exhibits erratic fluctuations as strain increases, which differs from previous theoretical conjectures. However, on average, we observe that the local shear stress of T1 events decreases from birth to death, consistent with trends observed in local pressure and the ratio of local shear stress to pressure.

We compare the average shear stress and pressure between T1 and non-T1 (NT1) particles and find that a significant drop in stresses occurs around $\gamma_y$, with comparable values before that. These findings establish a strong connection between microscopic and macroscopic plastic deformations. Additionally, we measure the spatial correlations of microscopic stress fluctuations, which exhibit a long-range anisotropic spatial correlation, similar in nature to the theoretically proposed Eshelby inclusion but with a $r^{-1.5}$ scaling. Remarkably, we observe similar correlation functions for T1 and NT1 regions, suggesting the existence of contact-scale plasticity in granular materials in addition to T1 plastic deformation.

\paragraph{Experimental Setup}
\begin{figure}
\includegraphics[width=8.6cm]{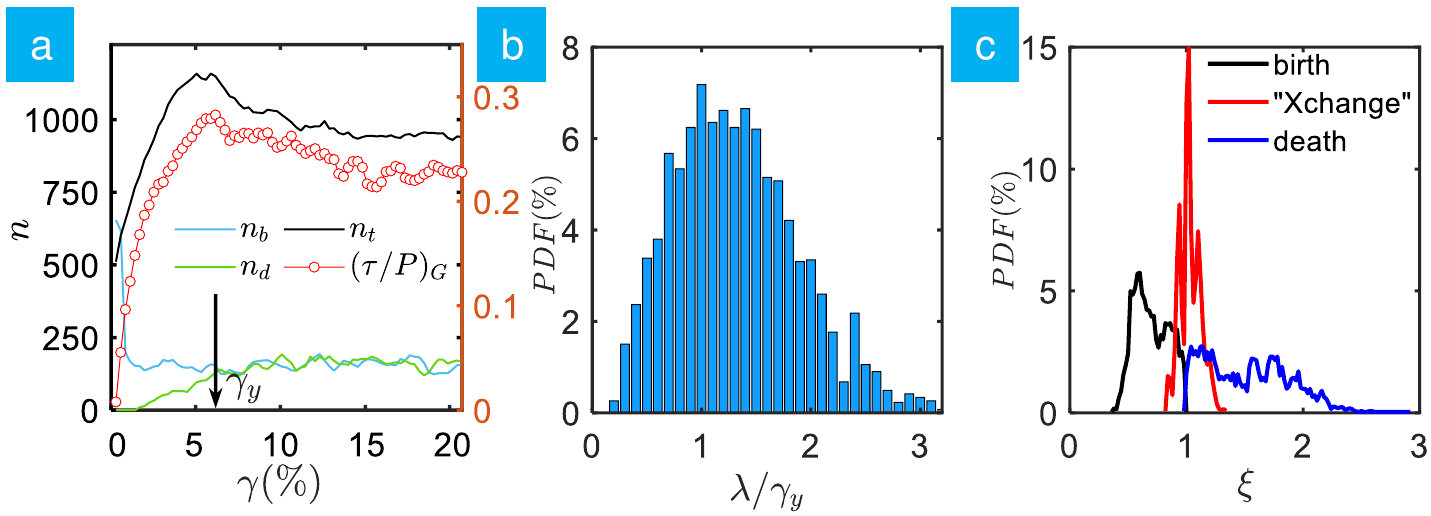}
\caption{(a) The double-Y plot draws the curves of the total ($n_t$), the birth ($n_b$), and the death ($n_d$) numbers of T1 events as a function of strain $\gamma$, and the corresponding ratio of the macroscopic shear stress over pressure $(\tau/P)_G$ of the system. (b) The PDF of T1 events’ lifetime $\lambda$. (c) The PDFs of $\xi$, which are related to the different statuses of T1’s birth, `Xchange' point, and death. The statistics in (c) are collected over different T1 events in space at a given strain $\gamma$ and over different strains $\gamma's$.}
\label{Fig2} 
\end{figure}

The experimental setup consists of a 2D pure shear system, as depicted in Fig.~\ref{Fig1}, following the configuration detailed in Ref.\cite{zheng}. A rectangular frame is subjected to compression along the x-axis and expansion along the y-axis under constant confining pressure, which is maintained using an air-bearing device. Inside the rectangular frame, photoelastic particles with diameters of 1 cm for large particles and 0.7 cm for small particles are immersed in density-matched brine to eliminate the base friction. Compression along the x-axis is applied incrementally with strain steps of $\delta\gamma=0.28\%$, with a maximum of 80 strain steps, starting from an isotropic jammed initial state. Particle configurations and stress images are recorded at every strain step using two Nikon cameras, as described in Ref.\cite{zheng}. Image processing techniques are employed to identify particle centers, inter-particle contacts, and track particle displacements over incremental strain steps\cite{zheng}. Inter-particle contact forces are measured with reasonable accuracy from the stress images\cite{photo_2,ling,YQWANG}, enabling the construction of particle-scale and macroscopic stress tensors\cite{zheng, jamming_nature,YQWANG}. To ensure statistical validity, multiple independent experimental runs are performed under the same protocol for ensemble averages. Confining boundary pressures ranging from 8.61 N/m to 10.04 N/m and 11.48 N/m are explored, confirming qualitatively similar results across different boundary pressures. Additionally, different system sizes ranging from 500, 2000, 5000, to 7000 disks are used to rule out finite size effects. Here, we present results only for a boundary pressure of 11.48 N/m and a system size of approximately 5000 disks with equal numbers of large and small disks. The results obtained for inter-particle friction coefficients of $\mu\approx0.1$ (corresponding to Teflon-wrapped disks) and $\mu\approx0.7$ (original disks) are qualitative similar. Therefore, we present results only for $\mu\approx0.1$.

\paragraph{Results}
Based on the critical roles of T1 events as elementary plastic units in the 2D bubble, foam and emulsion systems\cite{dennin_2004,foam_prl2,eric_emulsion,eric_emulsion2}, our goal is to identify T1 events in a 2D granular system under pure shear. To achieve this, we first perform Voronoi tessellation on the particle configuration at each strain step. This allows us to examine all clusters of Voronoi cells formed by four neighboring particles and identify the two nearest neighboring particles if they share a common edge. Subsequently, we track the topological changes of these four particles within each cluster, as T1 events are associated with topological transformations of the nearest neighbors, as illustrated in Fig.~\ref{Fig1}. (See the Supplementary Movies for the evolution of T1 events in an experimental run.) We define the topological state of this neighbor switching as the `Xchange' point of the T1 event.

The quantitative characterization of the topological exchange process in a T1 event can be achieved through the analysis of the aspect ratio $\xi$, which is defined as the ratio between the center-to-center distance of the nearest neighbor and that of the next nearest neighbors. An illustrative example of the evolution of $\xi$ for a T1 event is depicted in Fig.~\ref{Fig1}b, where $\xi$ remains nearly constant initially and then starts to rise monotonically within a certain range of strain intervals, eventually reaching a constant plateau as the global strain $\gamma$ increases further.

By disregarding small fluctuations of $\xi$ caused by mechanical noises, we can define the monotonic rising stage of $\xi$ as the effective evolution of the T1 event. As shown in Fig.\ref{Fig1}b, the starting ($\gamma_b$) and ending ($\gamma_d$) points of this rising region correspond to the birth and death points of the T1 event, respectively. Consequently, the strain interval (or lifetime) of the T1 event can be defined as $\lambda=\gamma_d-\gamma_b$. Fig.\ref{Fig1}a provides a typical snapshot of the spatial distribution of T1 events at $\gamma=2\gamma_y$, where $\gamma_y$ denotes the yielding strain. The color bar indicates the instantaneous values of $\xi$ for different T1 events. These T1 events are predominantly distributed in X-shaped shear bands, consistent with the distribution of nonaffine particle displacements (see the Supplementary Material for detail).

\begin{figure}[t]
\includegraphics[scale=.77]{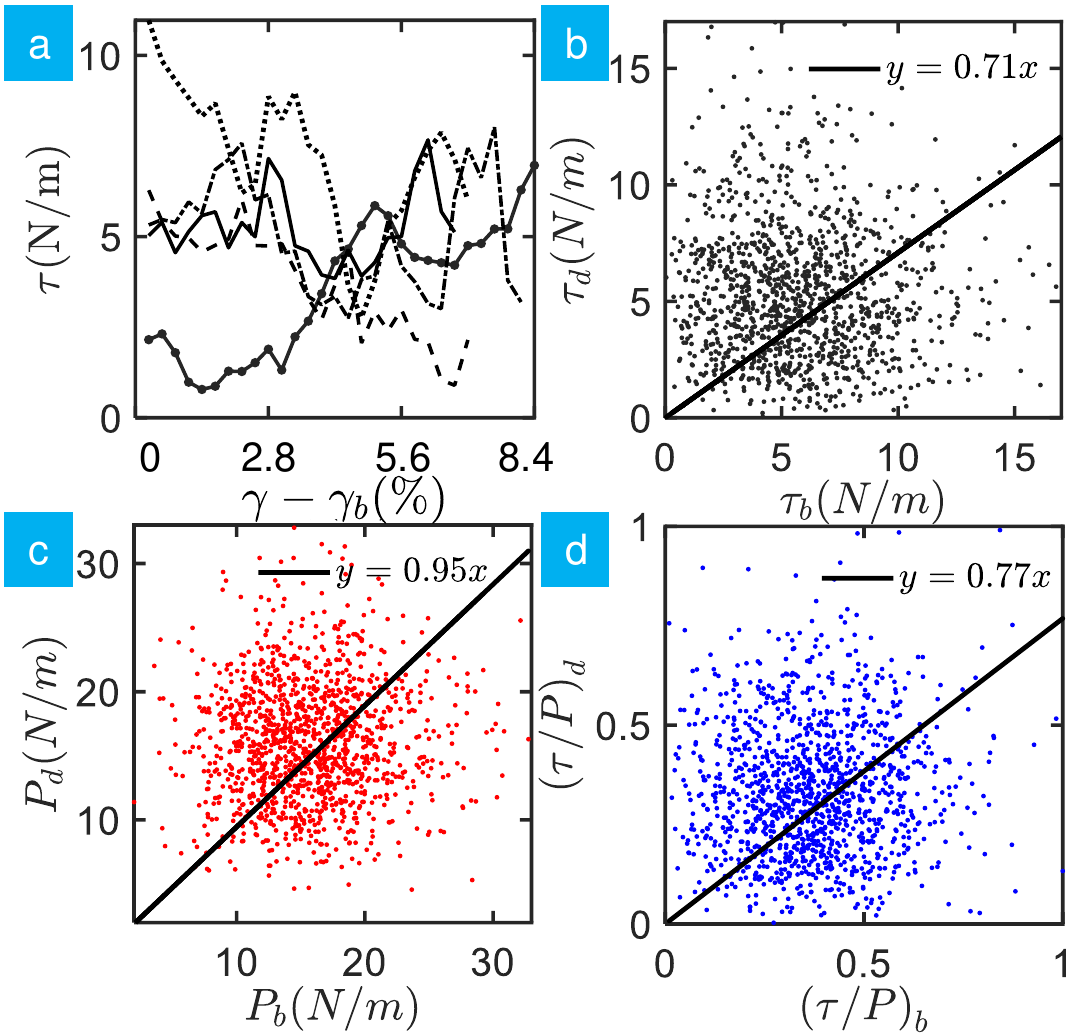}
\caption{(a) The transient evolution curves of local shear stress of several typical T1 events. (b-d) shows the scatter plots of birth (subscript b) vs death (subscript d) T1 particles’ local shear stress $\tau$ (panel b), local pressure $P$ (panel c) and local $\tau/P$ (panel c), respectively. The black line represents a linear fitting of all scattered points.
}
\label{Fig3} 
\end{figure}

Next, we investigate the relationship between the characteristics of T1 events and the macroscopic mechanical behavior of the system, as reflected by the ratio of the macroscopic shear stress over pressure $(\tau/P)_G$ curve (Fig.~\ref{Fig2}a). The curve of $(\tau/P)_G$ is plotted along with the curves of the numbers of birth ($n_b$), death ($n_d$), and total ($n_t=n_b+n_d$) T1 events. We observe that the trend of $(\tau/P)_G$ is similar to that of $n_t$, indicating the critical role of T1 events as microscopic plastic units. Notably, the curve of $n_b$ drops quickly to the steady state value, well before reaching $\gamma_y$, whereas the curve of $n_d$ slowly increases and eventually catches up with $n_b$. The balance between $n_b$ and $n_d$ occurs slightly after $\gamma_y$, where the $(\tau/P)_G$ curve peaks, defining the mechanical yield strain.
This interesting behavior is also reflected in the probability density distribution (PDF) of the lifetime $\lambda$ of T1 events, as shown in Fig.\ref{Fig2}b. The PDF peaks around $\lambda$, which is slightly larger than $\gamma_y$, although the distribution is broad. Fig.\ref{Fig2}c displays the PDFs of the aspect ratio $\xi$ of T1 events at the birth, death, and `Xchange' points. As expected, the distribution of the `Xchange' point exhibits a sharp peak around 1, indicating that `Xchange' events largely maintain the initial aspect ratio of the particles. In contrast, the PDFs of birth and death states show a relatively broad distribution range, indicating that T1 events can occur with different aspect ratios at the beginning and end. This behavior of T1 events is distinct from that of the bubble and emulsion systems \cite{dennin_2004,eric_emulsion,eric_emulsion2}, likely due to the stability of frictional particles and the athermal nature of the granular system\cite{bob_review,jamming_nature}.


\begin{figure}
\includegraphics[scale=.7]{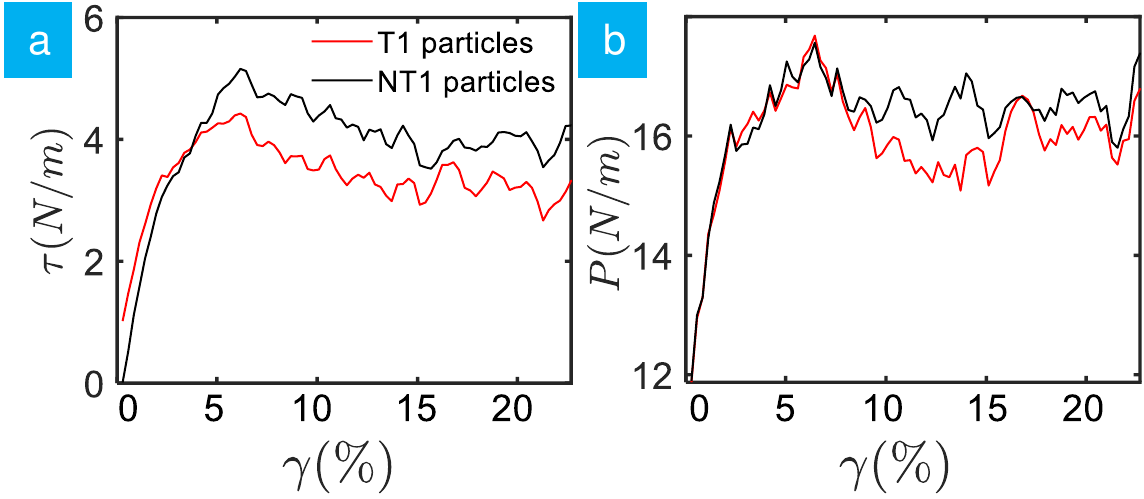}
\caption{(a) local shear stress $\tau$ and (b) local pressure $P$ as a function of strain $\gamma$.  Here $\tau$ and $P$ are ensemble averaged over all T1-event particles and non-T1-event particles at given $\gamma$, respectively.}
\label{Fig4} 
\end{figure}

\begin{figure}
\includegraphics[width=8.6cm]{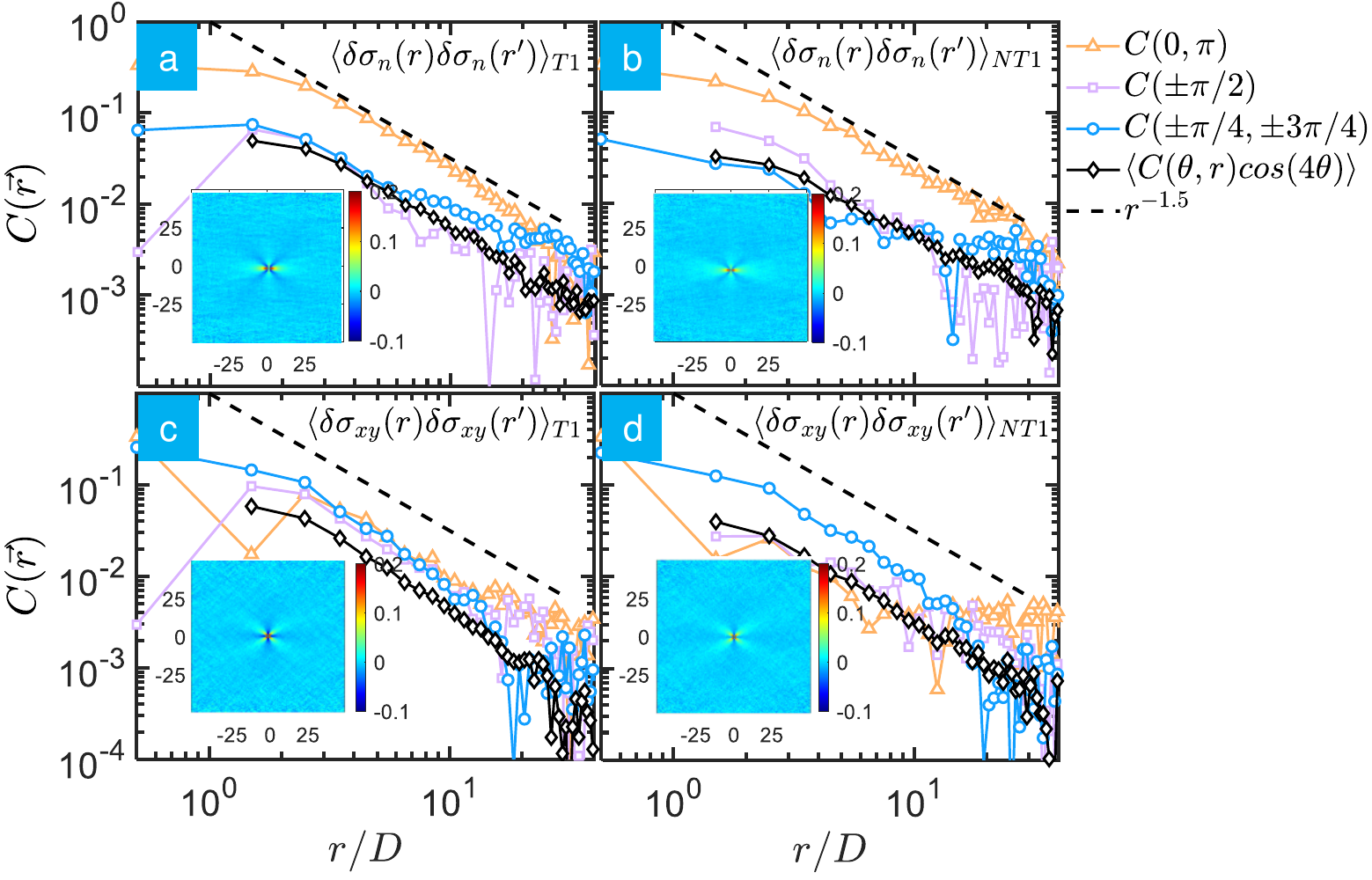}
\caption{(a-d) shows the spatial auto-correlation functions $C(r)$ of the local stress fluctuations of $\delta\sigma_n$ and $\delta\sigma_{xy}$ in the T1 and NT1 regions along different orientation angles, with the sub-panels showing the 2D correlation map $C(\vec{r})$, respectively. The axes of the map are rescaled by mean particle diameter. The legend indicates the correlation function along with different orientations.}
\label{Fig5} 
\end{figure}

It is believed that T1 events are responsible for local stress relaxations, leading to local stress drops \cite{langer_review,langer1998,langer2008}. This hypothesis was later confirmed in experiments with flowing emulsions\cite{eric_emulsion}. To investigate whether this phenomenon also occurs in granular materials, we need to analyze the micromechanics of T1 events.
Figure \ref{Fig3}a shows the evolution of local shear stress for several typical T1 events. The local stress of an individual T1 event, computed from the average over the stress tensors of the four particles involved, exhibits significant fluctuations with increasing strain, showing no discernible trend. However, when we compare the local stresses of all T1 events in the birth and death states using a scatter plot, we observe that, on average, the local shear stress $\tau_d$ in the death state is smaller than the local shear stress $\tau_b$ in the birth state, as shown in Fig.~\ref{Fig3}b. In contrast, the average drop in local pressure $P$ between the two states is much smaller, as depicted in Fig.~\ref{Fig3}c. As a result, the scatter plot of $\tau/P$ shows a comparable degree of average drop compared to the local shear stress, as shown in Fig.~\ref{Fig3}d. This suggests that, on average, shear stress is partially released while pressure remains nearly invariant after T1 events.
These results indicate that, on average, T1 events in granular materials are also responsible for local stress relaxations, similar to those observed in flowing emulsions\cite{eric_emulsion}. However, the changes in local stress for individual T1 events are much more complex, as evident from the large scattering of data points in Fig.~\ref{Fig3}.

To further investigate the micromechanical characteristics of T1 events, we compared the average stress values of T1 and NT1 particles at different strains $\gamma's$. Interestingly, as shown in Fig.~\ref{Fig4}a, for small $\gamma$ values, the average shear stress of T1 particles is slightly higher than that of NT1 particles. However, this trend is reversed when $\gamma$ exceeds a sufficiently large strain, with the average shear stress of T1 particles significantly dropping below that of NT1 particles. The crossing strain occurs slightly below $\gamma_y$.
In contrast, as seen in Fig.~\ref{Fig4}b, the average local pressure of T1 particles is comparable to that of NT1 particles before the yield strain, unlike the behavior of shear stress. But when $\gamma>\gamma_y$, the average pressure of T1 particles becomes significantly lower than that of NT1 particles.
Despite the quantitative differences in stress values between T1 and NT1 particles, the two curves in Fig.~\ref{Fig4} exhibit similar qualitative trends, suggesting a strong coupling in stress among the two sets of particles, possibly due to long-range correlations of stress.

In amorphous solids, it has been theorized for a long time that stress relaxations of local defects exhibit long-range correlations in space, akin to Eshelby inclusions in elastic materials \cite{EEshelby1957, argon, eshelby_pre2, procaccia_band}. This anisotropic response is believed to decay according to a power law. Previous measurements of local strain correlations in colloids\cite{weitz-science,weitz-pre,foam_eshelby} and local stress correlations in flowing emulsions\cite{eric_emulsion} have supported this conjecture. However, directly measuring spatial correlations of stress fluctuations in granular systems has been challenging and thus rare. 
We next present the spatial correlation functions $C(\vec{r})$ (whose definition is given in detail in the Supplementary Material) of stress fluctuations of $\delta\sigma_n$ and $\delta\sigma_{xy}$, as shown in Fig.~\ref{Fig5}. We have checked that the specific value of $\delta\gamma$ does not change the qualitative behaviors of results in Fig.~\ref{Fig5}. 
Remarkably, the 2D correlations displayed in Fig.~\ref{Fig5} exhibit the distinctive fourfold symmetry characteristic of Eshelby inclusions. However, the scaling of these correlations follows a novel power-law decay of $r^{-1.5}$, which is different from the expected $r^{-2}$ scaling of Eshelby inclusions\cite{EEshelby1957, eshelby_pre2}. Furthermore, surprisingly, the correlations of the T1 and NT1 regions are qualitatively and quantitatively indistinguishable. 
In the Supplementary Material, we also show that the auto-correlations of local strains in T1 and NT1 regions are qualitatively similar with each other, particularly for the prominent strain component $\epsilon_n$.

\paragraph{Discussion and Conclusion}
In this study, we conducted micromechanical measurements on densely packed bidisperse photoelastic disks under pure shear, with a focus on the statistics and dynamics of T1 events. We made an intriguing observation that the evolution of total T1 events closely resembles the stress-strain curve. The interplay between newly born and annihilated T1 events results in an initial balance point around the yield strain, followed by a dynamic equilibrium, which is consistent with existing theories\cite{langer_review, langer1998, langer2008}.

Despite the erratic fluctuations of local stresses in individual T1 events with respect to the global strain, the average trend supports the notion that T1 events can be considered as elementary plastic units occurring in the softer regions of the material. This is further corroborated by the average local stress drop associated with T1 events. Remarkably, the local stresses of T1 particles are significantly lower than those of non-T1 particles beyond the yield strain.

Furthermore, we made a noteworthy observation that the correlation functions of local stress fluctuations in both T1 and NT1 regions exhibit similar anisotropic power-law scaling of $\propto r^{-1.5}$, which suggests the presence of contact-scale plastic deformation as a distinct characteristic of granular materials, setting them apart from other amorphous solids.

\begin{acknowledgments}
This work is supported by the NSFC (No. 11974238, No. 11774221, and No. 12004241). The Innovation Program of Shanghai Municipal Education Commission under No. 2021-01-07-00-02-E00138 also supports this work. We also acknowledge the support from the Student Innovation Center of Shanghai Jiao Tong University.
\end{acknowledgments}

\appendix

\nocite{*}

\bibliography{MainText}

\end{document}



\title{Supplementary Material of Micromechanical measurements of local plastic events in granular materials}

\author{Jie Zheng}
 \affiliation{School of Physics and Astronomy, Shanghai Jiao Tong University}
\author{Aile Sun}
 \affiliation{School of Physics and Astronomy, Shanghai Jiao Tong University}
\author{Jie Zhang}%
 \email{jiezhang2012@sjtu.edu.cn}
\affiliation{School of Physics and Astronomy, Shanghai Jiao Tong University}%
\affiliation{Institute of Natural Sciences, Shanghai Jiao Tong University}%



\date{\today}

\maketitle

\subsection{1. Comparison of T1 events vs nonaffine motion}
Fig.~\ref{SFig1} compares the spatial distribution of T1 events at steady state shear with the distribution of particles with the large nonaffine motion [1]. Note, for comparison, we select the average lifetime of T1 events in Fig.2b of the main text as the strain interval to calculate nonaffine motion value $D_{min}^2$ [1].

\begin{figure*}
\centering
\includegraphics[width=0.9\textwidth]{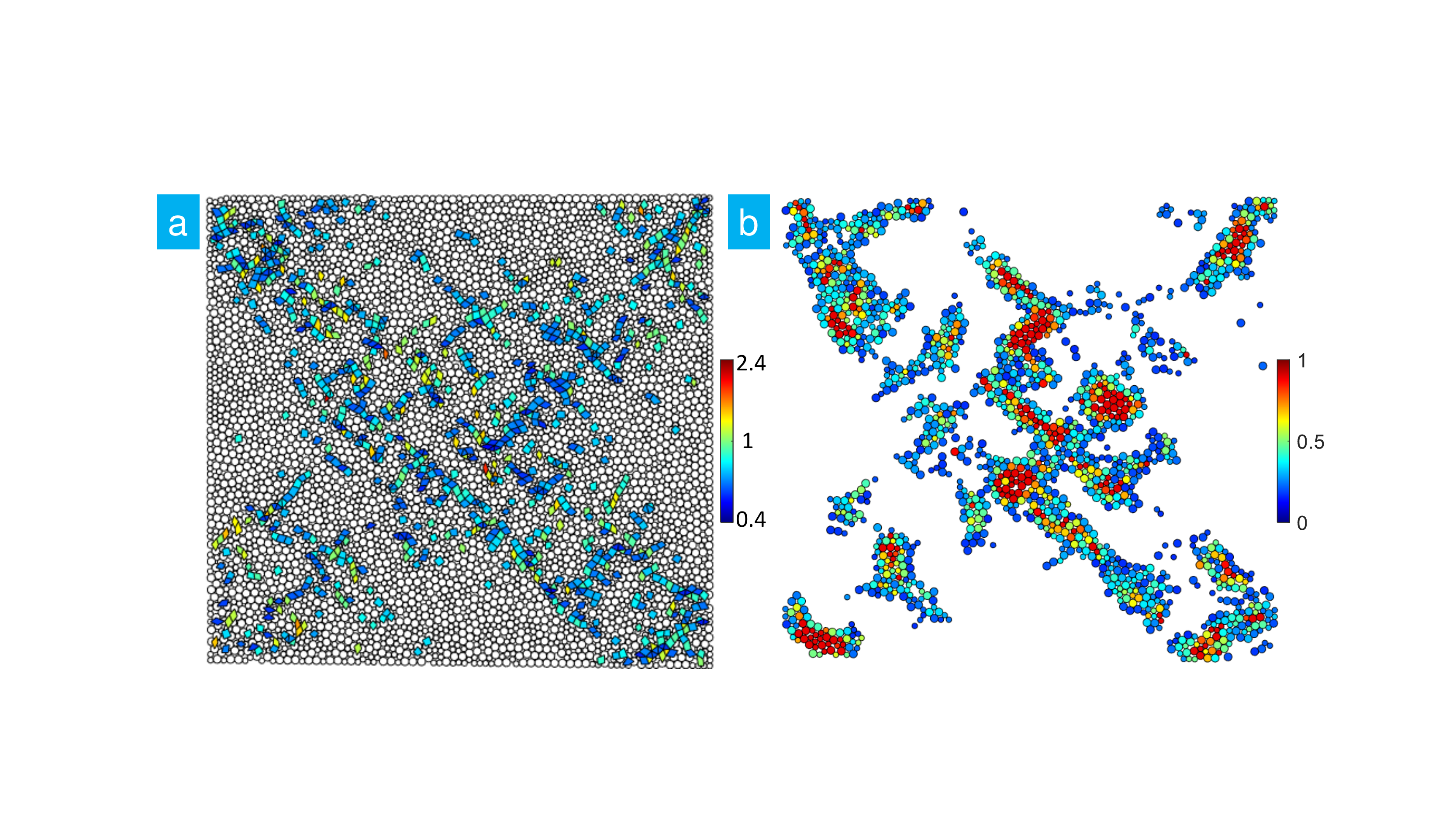}
\caption{\label{SFig1}(a) A typical spatial distribution of the T1 events (quadrilaterals). The filled colors represent the instantaneous values of aspect ratios $\xi$ of individual T1 events. (b) the spatial distribution of particles with large $D_{min}^2$, with the filled colors labels the relative nonaffine value. Only the particles of top $30\%$ nonaffine values are shown for better visualization.}
\label{SFig1} 
\end{figure*}

\begin{figure*}
\centering
\includegraphics[width=1.8\textwidth]{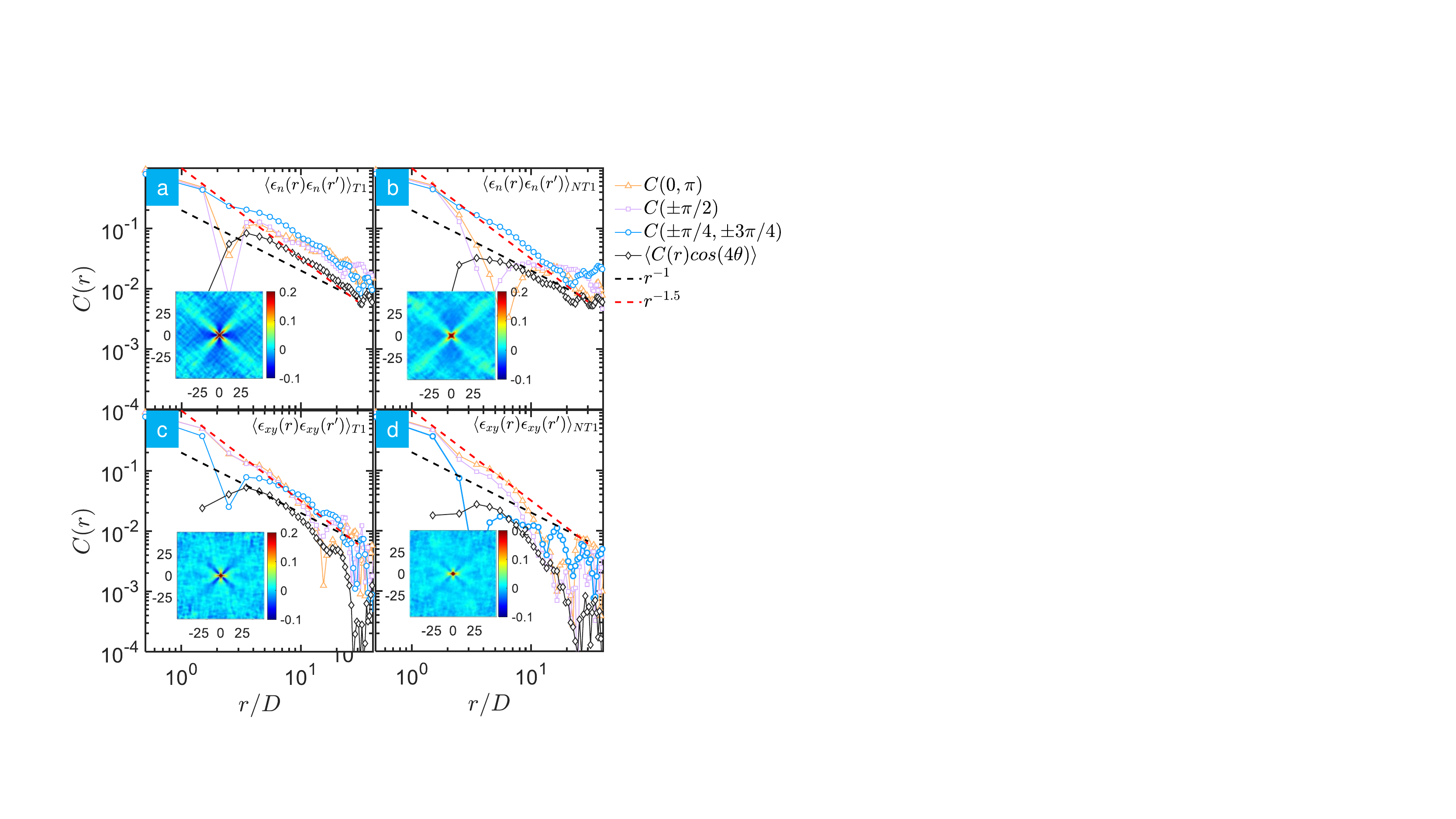}
\caption{\label{SFig2}(a-d) show the spatial auto-correlation functions of the local stress fluctuations of $\delta\sigma_n$ and $\delta\sigma_{xy}$ in the T1 and NT1 regions, with the insets showing the two-dimensional correlation maps $C(\boldsymbol{r})$. The horizontal axis of each map is rescaled by mean particle diameter, and the legend indicates the correlation functions along different orientations.}
\label{SFig2} 
\end{figure*}

\subsection{2. The spatial auto-correlations in the T1 and NT1 regions}

 We calculate the two-dimension auto-correlation map $C(\vec{r})$ according the formula of
 
 $C(\vec{r})=\frac{1}{\Delta A}\langle A(\boldsymbol(\vec{r}+\vec{r_i}))A(\boldsymbol(\vec{r_i})  \rangle_{\vec{r_i}}$,
 
 with $\langle ...\rangle _{\vec{r_i}}$ representing the statistical average covering all the T1 or NT1 particles, respectively. $\Delta A$ is the corresponding normalization factor. The variable $A$ refers to the local strains or stress fluctuations that need to be calculated, etc. 
 For example, in the correlations of stress fluctuations, the variable A refers to $\delta\sigma_n$ and $\delta\sigma_{xy}$ respectively. Here $\delta\sigma_n=\sigma_n(\gamma+\delta\gamma)-\sigma_n(\gamma)$, with $\sigma_n=\sigma_{xx}-\sigma_{yy}$, and the definition of $\delta\sigma_{xy}$ is similar, with $\sigma_{xy}=(\sigma_{xy}+\sigma_{yx})/2$.
 Note that for better statistics, all the $C(\vec{r})$ calculation is ensemble averaged over all the $\gamma's$ of 10 realizations. We have checked that the specific strain interval used to calculate the variable $A$ does not change the qualitative results in this work.

\subsection{3. Comparisons between T1 and NT1 particles' spatial auto-correlations of local strain}
Fig.~\ref{SFig2} shows the spatial auto-correlation functions of local strains $\epsilon_n$ and $\epsilon_{xy}$ in the regions where T1 and NT1 particles are respectively located. 
Here, by definition, the local strain $\epsilon_n=\epsilon_{xx}-\epsilon_{yy}$, and $\epsilon_{xy}=(\epsilon_{xy}+\epsilon_{yx})/2$.
Being analogous to the local stress fluctuations, as shown in the main text, the fourfold feature of the Eshelby-like correlation map is very prominent. The power law behaviors of correlation functions in T1 local regions seem more convincing than in NT1 regions in $\epsilon_{xy}$, possibly due to the relatively weaker signal of $\epsilon_{xy}$ compared to $\epsilon_n$ in pure shear.

\appendix

\nocite{*}


\subsection{Reference}
[1] M. L. Falk and J.S. Langer, Dynamics of viscoplastic deformation in amorphous solids, Phys. Rev. E. 1998